\documentclass[a4paper,fleqn]{cas-sc}
\usepackage{soul}
\usepackage[sort&compress,numbers]{natbib}
\usepackage{lineno}
\usepackage{setspace}
\usepackage{verbatim}
\usepackage{graphics}
\usepackage{graphicx}
\usepackage{cuted}
\usepackage{lipsum}
\usepackage{multirow}

\def\tsc#1{\csdef{#1}{\textsc{\lowercase{#1}}\xspace}}
\tsc{WGM}
\tsc{QE}
\tsc{EP}
\tsc{PMS}
\tsc{BEC}
\tsc{DE}

\begin{document}

\let\WriteBookmarks\relax
\def\floatpagepagefraction{1}
\def\textpagefraction{.001}
\shorttitle{Simulating Elasticity and Fracture in Me-Graphene Monolayers and Nanotubes}
\shortauthors{Pereira Jr \textit{et~al}.}

\title [mode = title]{Exploring the Elastic Properties and Fracture Patterns of Me-Graphene Monolayers and Nanotubes through Reactive Molecular Dynamics Simulations}

\author[1]{Marcelo L. Pereira Junior}
\cormark[1]
\ead{marcelo.lopes@unb.br}
\author[2,3]{Jos\'e. M. De Sousa}
\author[4]{Wjefferson H. S. Brand\~ao}
 \author[3,5]{Douglas. S. Galv\~ao}
\author[3]{Alexandre F. Fonseca}
 \author[6]{Luiz A. Ribeiro Junior}

 \address[1]{University of Bras\'{i}lia, Faculty of Technology, Department of Electrical Engineering, Bras\'{i}lia, Brazil.}
\address[2]{Federal Institute of Education, Science and Technology of Piau\'i -- IFPI, S\~ao Raimundo Notato, 64770-000, Piau\'i, Brazil.}
\address[3]{Applied Physics Department, ``Gleb Wataghin'' Institute of Physics - IFGW, University of Campinas -- UNICAMP, Rua Sérgio Buarque de Holanda, 777 - Cidade Universitária, 13083-859, Campinas, S\~ao Paulo, Brazil.}
\address[4]{Department of Physics, Federal University of Piauí, Terezina, Piauí, Brazil.}
\address[5]{Center for Computing in Engineering \& Sciences, Unicamp, Campinas, SP, Brazil.}
\address[6]{University of Bras\'{i}lia, Institute of Physics, Bras\'{i}lia, Brazil.}

\begin{abstract}
Me-graphene (MeG) is a novel two-dimensional (2D) carbon allotrope. Due to its attractive electronic and structural properties, it is important to study the mechanical behavior of MeG in its monolayer and nanotube topologies. In this work, we conducted fully atomistic reactive molecular dynamics simulations using the Tersoff force field to investigate their mechanical properties and fracture patterns. Our results indicate that Young's modulus of MeG monolayers is about 414 GPa and in the range of 421-483 GPa for the nanotubes investigated here. MeG monolayers and MeGNTs directly undergo from elastic to complete fracture under critical strain without a plastic regime.
\end{abstract}




\maketitle
\doublespacing

\section{Introduction}

Nanoscience is a rapidly growing area with great potential for developing new materials \cite{whitesides2005nanoscience,hutchison2008greener,braun1997nanoscience}. The experimental realization of graphene \cite{novoselov2004electric} has been a breakthrough. It has since been extensively studied in nanoelectronics due to its lightweight nature, high mechanical strength, good transparency, and excellent electrical and thermal conductivity \cite{geim2009graphene,lui2009ultraflat,XuPereira2014}. This achievement has inspired numerous experimental \cite{sundqvist2021carbon,tang2014two} and theoretical \cite{xu2014two,fan2017new} investigations on developing new 2D carbon allotropes. The primary objective is to create all-carbon 2D materials sharing some of graphene's exceptional properties, which could overcome its limitations in flat electronics, particularly its lack of an electronic bandgap \cite{withers2010electron}.

Carbon allotropes such as graphite \cite{chung1987exfoliation}, carbon nanotubes \cite{iijima1991helical}, and graphene \cite{novoselov2004electric} are mainly composed of sp$^2$-hybridized carbon atoms forming hexagonal rings, which provide excellent structural stability. In contrast, pentagonal rings are less commonly found in stable all-carbon materials such as C$_{60}$ molecules \cite{kroto1985c60}, primarily due to unfavorable bonding angles that may lead to structural instability. However, 2D carbon allotropes with pentagonal rings can exhibit semiconductor behavior \cite{zhang2015penta}, and the search for new stable carbon structures containing pentagons is now attracting much interest \cite{jana2021emerging}.

A novel 2D carbon allotrope, similar to graphene, has recently been proposed and named Me-graphene (MeG) \cite{zhuo2020me}. It comprises 5-6-8 rings of carbon atoms with sp$^2$ and sp$^3$ hybridizations, resulting in a semiconductor structure. Topologically, MeG can be formed by assembling C$-$(C$_3$H$_2$)$_4$ molecules. It exhibits a buckled layer but no negative frequencies in the phonon spectra, suggesting its structural stability. MeG exhibits properties between graphene and penta-graphene in terms of energy formation and non-null electronic bandgap. It also presents a near-zero Poisson's ratio behavior over a wide range of strain values. It has an indirect band gap of 2.04 eV, which evolves to a direct band gap of 2.62 eV under compressive strain. Furthermore, it is predicted a high hole mobility of 1.60$\times10^5$ cm$^2$V$^{-1}$s$^{-1}$ at 300 K. Since MeG presents electronic and structural features desirable in flat nanoelectronics applications, it is important to investigate its mechanical behavior \cite{zhuo2020me,luo2023carbon}, for both monolayer and nanotube forms, to improve our understanding of these family of nanostructures.

In the present work, we have used full atomistic reactive classical molecular dynamics (MD) simulations to investigate the mechanical properties and fracture patterns of MeG monolayers (MeGM) and nanotubes (MeGNTs), see Figure \ref{fig1}. Our simulations considered different tube diameters at room temperature. Young's modulus values of MeG monolayers were estimated to be about 414 GPa and can vary between 421-483 GPa for the nanotubes considered here. Upon reaching a critical strain, MeGM undergoes complete fracture without exhibiting a plastic deformation regime (brittle behavior). MeGNTs, on the other hand, exhibit a flat plastic region between the elastic and the completely fractured regimes.

\section{Methodology}

We performed MD Simulations using the LAMMPS \cite{plimpton1995fast} code and the Tersoff force field  \cite{tersoff1989modeling}. Tersoff goes beyond non-reactive force fields, allowing the description of forming and breaking bonds, which is necessary to investigate non-linear structural deformations and fracture dynamics. 

\begin{figure}[pos = htb]
	\centering
	\includegraphics[width=\linewidth]{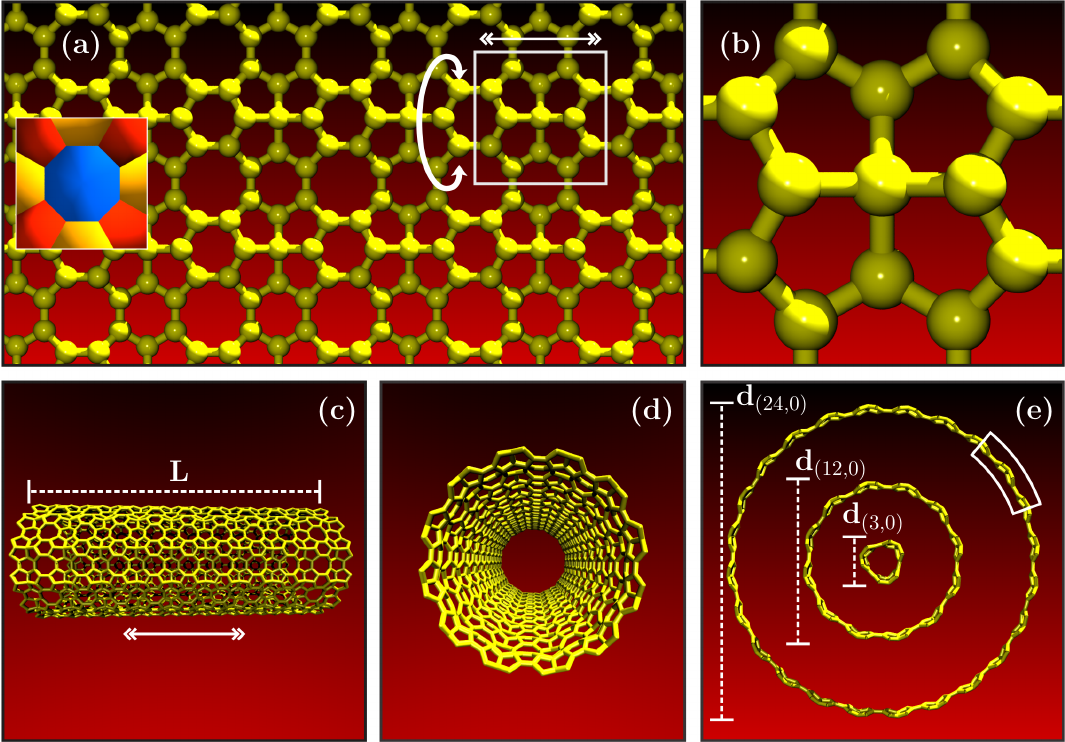}
	\caption{Schematic representation of (a-b) Me-graphene monolayers (MeGMs) and (c-e) Me-graphene nanotubes (MeGNTs) topologies. The MeGNTs studied here have the same length (\textbf{L}) but distinct diameters (\textbf{d}). See text for discussions.}
	\label{fig1}
\end{figure}

In Figure \ref{fig1}(a), we present the MeGM structure, where its lattice unit cell (magnified in Figure \ref{fig1}(b)) is indicated by a white square and the rolling direction used to generate the MeGNTS by a curvy arrow. The MeG carbon atoms forming the different rings are indicated in the colored box (red, blue, and yellow for fused pentagons, octagons, and hexagons). The `PaperChain' representation implemented in the molecular visual dynamics (VMD) \cite{humphrey1996vmd} software was used to render the ring composition. This technique finds all rings up to a user-defined maximum size by analyzing the molecular topology and fitting a polyhedron to the selected atoms and the ring centroid \cite{humphrey1996vmd}. 

In Figure \ref{fig1}(c), we present a side view of the MeGNTs, with their length indicated by \textbf{L}. Figures \ref{fig1}(d) and \ref{fig1}(e) illustrate the front views of MeGNTs of different diameters (\textbf{d}). The white rectangle in Figure \ref{fig1}(e) highlights MeGMs and MeGNTs buckled morphologies. According to the study by Luo et al. \cite{luo2023carbon}, MeG exhibits a blocking of 1.01 \r{A}. Consequently, the thickness of both MeGM and MeGNT is reported to be 4.41 \r{A} \cite{luo2023carbon}. The MD snapshots were rendered using VMD software, while the Sculptor \cite{humphrey_JMG} software suite was used to generate the MeGNT structures.

The used MeGM structure models investigated here are composed of an 18x18 replication of the MeG unit cell ($5.74\times5.74$ \r{A}$^{2}$ with 13 carbon atoms), resulting in a membrane with dimensions of $103.3\times103.3$ \r{A}$^{2}$ and containing a total of 4212 carbon atoms. The MeGNT(3,0), MeGNT(12,0), and MeGNT(24,0) structures were generated from $3\times18$ (\textbf{d}$_{(3,0)}=5.5$ \r{A}), $12\times18$ (\textbf{d}$_{(12,0)}=21.9$ \r{A}), and $24\times18$ (\textbf{d}$_{(24,0)}= 43.9$ \r{A}) supercells, respectively. The length of all MeGNTs is the same and equals 103.3 \r{A}. The total number of carbon atoms in MeGNT(3,0), MeGNT(12,0), and MeGNT(24,0) are 702, 2808, and 5616, respectively. We applied periodic boundary conditions to both systems. In the case of MeG, the periodicity encompasses the x and y directions of the plane. However, for MeGNTs, the periodicity is limited to the longitudinal z-direction only.

MeGNTs are labeled using integer indices similar to those for graphene nanotubes \cite{samsonidze2003concept}, i.e., MeGNT$(m,n)$. $m$ and $n$ are used linearly to determine the nanotube circumference \cite{luo2023carbon}. The MeG unit cell has symmetric orthogonal lattice vectors so that $a$ and $b$ are equal. Therefore, MeGNT$(m,0)$ and MeGNT$(0,m)$ nanotubes are identical. As mentioned above, we have considered the MeGNT(3,0), MeGNT(12,0), and MeGNT(24,0) structures. The diameter of each tube was calculated using \textbf{d}$_{(m,0)} = ma / \pi$, where $a$ is the lattice constant.

We used the velocity Verlet algorithm with a time step of 0.05 fs \cite{martys1999velocity} to integrate Newton's equation of motion. Before carrying out the tensile loading simulations on our model nanostructures, we thermalized them using a Nos\'e-Hoover thermostat for 50 ps \cite{evans1985nose}. Afterward, an NPT simulation with zero pressure was performed for an additional 50 ps to eliminate residual lattice stress in the thermalized systems \cite{andersen1980molecular}.

To obtain the elastic properties and fracture patterns at 300 K, we stretched MeGM and MeGNTs using a constant engineering tensile strain rate of $10^{-6}/$fs using NPT and NVT ensembles, respectively. The NPT ensemble is typically preferred for 2D and 3D systems under strain as it allows for adjustments in the system volume. On the other hand, the NVT ensemble is generally adopted for 1D systems since only one of the box directions was undergoing deformation. The mechanical properties of MeGM and MeGNTs were studied by applying stresses to these nanostructures along the x-direction (layers) and z-direction (tubes), respectively, while the other directions can relax freely. The stretching processes are maintained until the complete structural failure (fracture).

\section{Results}

\begin{figure}[pos = htb!]
\begin{center}
\includegraphics[width=0.6\linewidth]{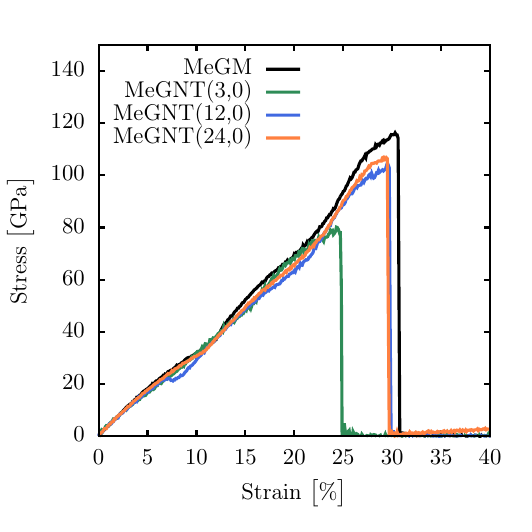}
\caption{Stress-strain curves for MeGM and MeGNTs as a function of the applied strain.}
\label{fig2}
\end{center}
\end{figure}

We begin our discussion by presenting the stress-strain interplay in MeGM and MeGNT systems. Figure \ref{fig2} depicts the stress-strain curves for MeGM and MeGNTS at 300 K, with the stretching tensile loading applied along the x- and z-directions. The stress-strain curves show that these systems directly undergo from elastic to a completely fractured regime (zero stress regions) at a critical strain. We can observe that the stress-strain relationship exhibits qualitative similarities between the two systems. However, there is a minor yet noticeable difference for MeGNT(3,0). This discrepancy arises due to the tube's small diameter and pronounced curvature, which lead to its collapse during the tensile stretching process.  

In Table \ref{tab1}, we present a summary of some of the estimated elastic parameters extracted from the stress-strain curves of Figure \ref{fig2}. The $Y_M$ denotes Young's modulus values. The $\sigma_C$ is the critical stress value. $\epsilon_C$ denotes the value at the moment of the first fracture. We calculated the virial stress $\sigma_C$ along the stretching direction according to the approach described in reference \cite{de2021nanostructuresgraphene}. Except for the atypical MeGNT(3,0) case, MeGNTs generally exhibit similar values for $Y_M$ and critical strains compared to MeGM. As the diameter of the tubes increases beyond MeGNT(12,0), the decrease in Young's modulus implies that the impact of curvature effects becomes less significant. These findings agree with the ones reported using density functional theory (DFT) calculations \cite{luo2023carbon}. 

Previous studies \cite{luo2023carbon,zhuo2020me} have investigated the mechanical properties of MeGM and MeGNTs using DFT calculations. These studies reported $Y_M$ values of approximately 477 GPa for MeGM and 478-551 GPa for MeGNTs. For increasing diameter values, the $Y_M$ of MeGNTs is comparable to MeGM's \cite{luo2023carbon}. It is important to note that the DFT calculations were performed at zero temperature, where the dominant factor influencing the $Y_M$ values of MeGNTs is the curvature of the nanotube. As shown in Table \ref{tab1}, temperature slightly affects the estimated $Y_M$ values when contrasting MD and DFT values. Despite differences in the approach used to estimate $Y_M$ between the referenced study \cite{luo2023carbon} and our current study, we still observe a consistent trend where the $Y_M$ values of MeGNTs follow a similar pattern to those of MeGM as the diameters increase. Moreover, there is a good agreement between the elastic properties obtained using MD and DFT simulations.

As mentioned above, the MeGNTs stand for strain and stress values comparable to MeG before breaking than the MeGM. Their critical stress values are slightly lower than the one for graphene (about 130 GPa (25\%) \cite{lee2008measurement}), but in the same range as other 2D all-carbon allotropes theoretically proposed \cite{junior2023irida,junior2020elastic,sui2017morphology,sun2016new}. The critical stress values for the MeGNTs are almost the same for the different diameters considered here, while the critical strain is slightly diameter dependent (see Table \ref{tab1}), thus indicating that one of the curvature effects on the structure is making them more compliant. 

\begin{table}[pos = htb!]
\centering

\begin{tabular}{|c|c|c|c|}
\hline
System      & $Y_M$  (GPa)                 & $\sigma_C$ (GPa) & $\epsilon_C$ (\%) \\ \hline
MeGM        & 414.5 & 116.2        & 30.0           \\ \hline
MeGNT(3,0)  & 483.4 & 80.1        & 24.3           \\ \hline
MeGNT(12,0) & 422.8 & 103.8        & 29.6           \\ \hline
MeGNT(24,0) & 421.2 & 106.9        & 29.1           \\ \hline
\end{tabular}

\label{tab1}
\caption{Elastic properties for the MeGM and MeGNTs estimated from the stress-strain curves shown in Figure \ref{fig2}.}
\end{table}

\begin{figure}[pos = htb!]
\begin{center}
\includegraphics[width=\linewidth]{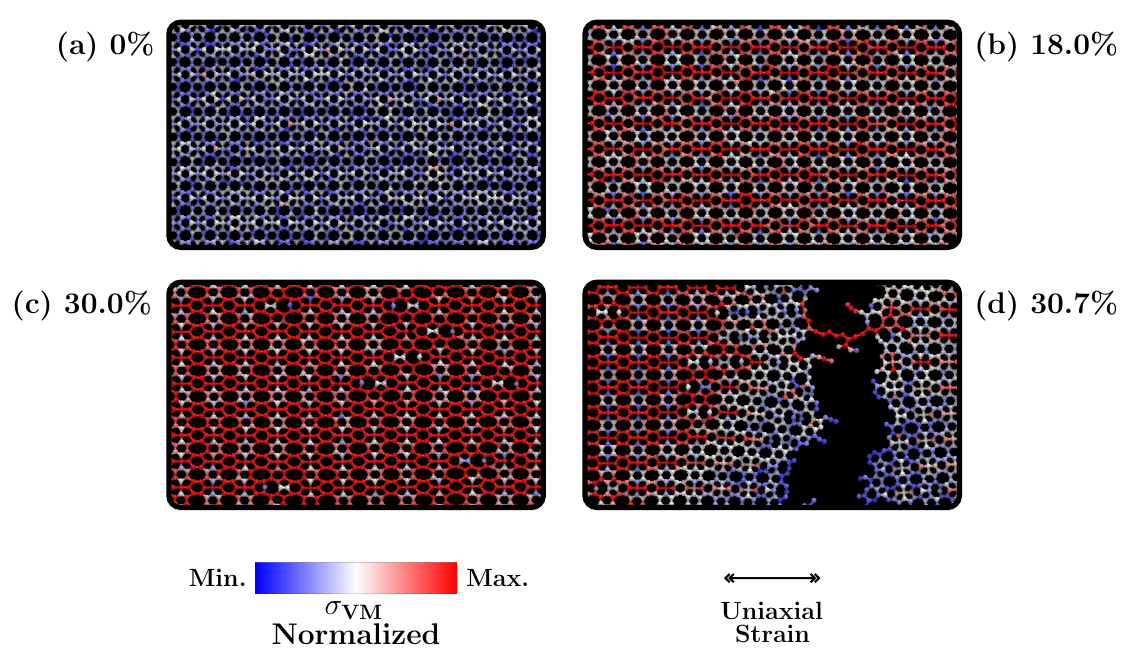}
\caption{Representative MD snapshots for the fracture process of MeGM at 300K, stressed along the x-direction. Panels (a), (b), (c), and (d) show the MeGM bond arrangements when it is subjected to 0.0\%, 18.0\%, 30.0\%, and 30.7\% of strain, respectively.}
\label{fig3}
\end{center}
\end{figure}

Next, we examined the fracture patterns of MeGM obtained from the MD simulations. In Figure \ref{fig3}, we present typical MD snapshots of the fracture process for MeGM under x-directional stress at 300 K. The color scheme used in this Figure represents the von Mises (VM) stress per-atom values \cite{von1913mechanics}, with red and blue colors indicating high- and low-stress accumulation, respectively. These VM values can better identify the point or regions where the fracture starts and/or propagates. Further information about VM calculations can be found in references \cite{simulator2012lammps,brandao2021atomistic}.

Figures \ref{fig3}(a) and \ref{fig3}(b) show the MeGM lattice at 0\% and 18.0\% of strain, respectively. In these scenarios, the MeG structure has no failure. At a critical strain of 30.0\%, the first bond breaking occurs (see Figure \ref{fig3}(c)). Up to 30.7\% of strain, MeGM undergoes an abrupt transition from an elastic to a completely fractured regime, as shown in Figure \ref{fig3}(d). The deformation of the structure is entirely elastic, meaning that it could return to its original configuration when released from that strain without forming any defects. At a slightly higher strain (30.7\%, Figure \ref{fig3}(d)), several additional bonds break, and linear carbon chain (LAC) formations occur. The Supplementary Material presents an MD video for the stretching simulation of the MeG membrane.

The MeG structure is wholly fractured by less than 0.1\% of strain above the critical one, justifying this transition to be considered abrupt. The fracture occurs due to the breaking of two parallel bonds connecting a six-membered atomic ring with two neighboring rings composed of eight atoms, as illustrated in Figures \ref{fig3}(c-d). The crack then rapidly propagates along the y-direction, opposite to the direction of tensile loading, as shown in Figure \ref{fig3}(d). The MeGM fractures almost into two symmetrical unstressed moieties at the end of the stretching process.

\begin{figure}[pos = htb!]
\begin{center}
\includegraphics[width=\linewidth]{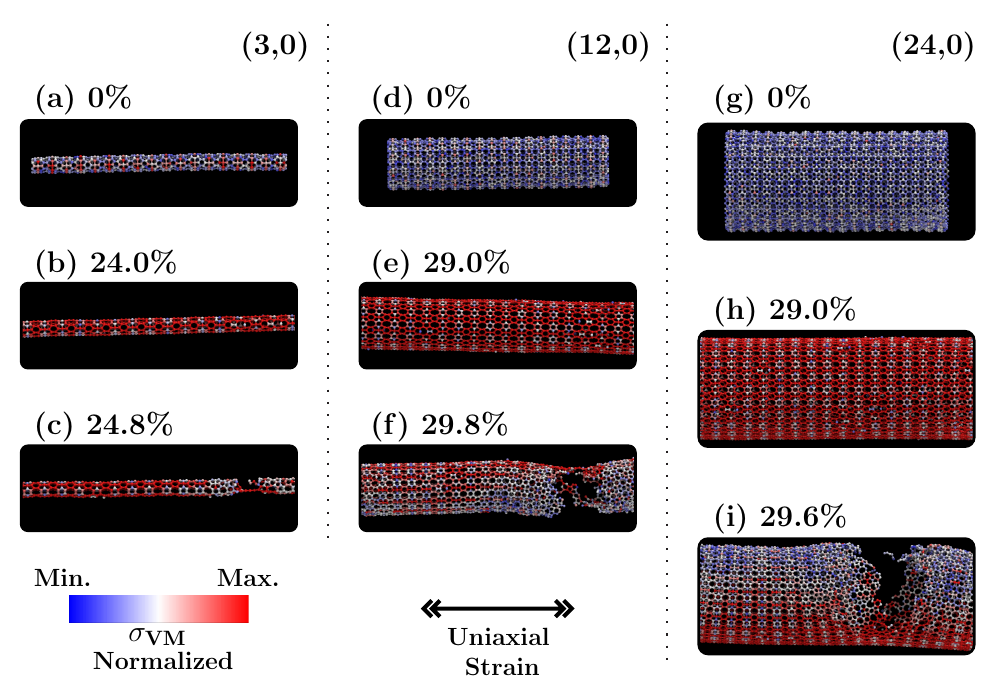}
\caption{Representative MD snapshots for the fracture process of (a-c) MeGNT(3,0), (d-f) MeGNT(12,0), and (g-i) MeGNT(24,0) at 300K, stressed along their axial direction.}
\label{fig4}
\end{center}
\end{figure} 

Finally, we show representative MD snapshots for the fracture process of the MeGNTs studied here. Again, we use a color scheme representing the VM stress per-atom values to help visualization of the fracture process. For MeGNT(3,0), the critical strain is 24.0\% (see Figures \ref{fig4}(a-c)). The Supplementary Material presents the MD videos for all the MeGNT cases discussed here. Figure \ref{fig4}(b) shows the MeGNT(3,0) structure at the critical strain, with all rings undergoing deformations along the longitudinal direction of the tube. The breaking bonds align with the direction of the applied strain. The MeGNT(3,0) fragmentation is abrupt, occurs at 24.8\% of strain, and also creates LACS resulting in two moieties, as shown in Figure \ref{fig4}(c). 

Similar fracture patterns were observed for MeGNT(12,0) and MeGNT(24,0). Figures \ref{fig4}(d) and \ref{fig4}(g) show the morphologies of these nanotubes for 0\% of strain. The first bond breakings occur for strain values up to 29.0\% (Figures \ref{fig4}(e) and \ref{fig4}(h)) of strain for MeGNT(12,0) and MeGNT(24,0), respectively. Subsequently, MeGNT(12,0) separates into two parts at 29.8\% (Figure \ref{fig4}(f)), whereas MeGNT(24,0) fractures by showing LACs at 29.6\% (Figure \ref{fig4}(i)) of strain. This nanotube is separated into two parts connected by only one LAC for strain values higher than 30\% (see Supplementary Material). 

The stress accumulation observed in the case of nanotubes, particularly for MeGNT(3,0), is due to the curvature of the system. Greater curvature corresponds to an increased tendency for stress accumulation. The curvature leads to stress concentrations that persist even after equilibration. Notably, the stress tends to accumulate in the bonds formed by the fused pentagons, the stiffest regions of the MeG structure.

\section{Conclusion}

In summary, we carried out extensive fully atomistic reactive molecular dynamics simulations to investigate the mechanical properties and fracture patterns of Me-graphene monolayer (MeGM) and ant their nanotubes (MeGNTs) under different temperature regimes and considering tubes of different diameters. We used the Tersoff potential \cite{tersoff1989modeling} for describing the forming and breaking bonds, which is necessary to investigate non-linear structural deformations and fracture dynamics. 

Our results reveal that the MeGM Young's modulus value is approximately 414 GPa, while the corresponding ones for the MeGNTs are in the range of 421-483 GPa. These systems abruptly undergo from elastic to completely fractured regimes (without a plastic regime). Additionally, the crack propagation for all structures occurs opposite to the applied tensile loading. The MeGM and MeGNT critical stress values are slightly smaller than the corresponding graphene one (about 116 GPa) but in the same range as other 2D all-carbon allotropes theoretically proposed in the literature. 

Previous DFT-based studies have reported YM values of approximately 477 GPa for MeGM and 478-551 GPa for MeGNTs \cite{luo2023carbon}. The $Y_M$ of MeGNTs for increasing diameter values is comparable to MeGM's. These DFT calculations were performed at zero temperature, where the dominant factor influencing the $Y_M$ values of MeGNTs is the curvature of the nanotube. However, in MD simulations, the system temperature also affects the estimated $Y_M$ values. Despite differences in the DFT and MD approaches to estimating $Y_M$, we still observe a consistent trend where the $Y_M$ values of MeGNTs follow a similar pattern to those of MeGM as the diameters increase. These findings agree with the ones reported using DFT calculations \cite{luo2023carbon}. 

\section*{Acknowledgements}

This work was financed in part by the Coordenaç\~ao de Aperfeiçoamento de Pessoal de Nível Superior (CAPES) - Finance Code 001 and grant 88887.691997/2022-00, Conselho Nacional de Desenvolvimento Científico e Tecnol\'ogico (CNPq), FAP-DF, FAPESP and FAPEPI. We thank the Center for Computing in Engineering and Sciences at Unicamp for financial support through the FAPESP/CEPID Grants \#2013/08293-7 and \#2018/11352-7. L.A.R.J acknowledges the financial support from FAP-DF grants $00193-00000857/2021-14$, $00193-00000853/2021-28$, $00193-00000811/2021-97$, and $00193.00001808/2022-71$, FAPDF-PRONEM grant $00193.00001247/2021-20$, and CNPq grant $302922/2021-0$. L.A.R.J. acknowledges N\'ucleo de Computaç\~ao de Alto Desempenho (NACAD) and for providing the computational facilities. A.F.F. thanks the Brazilian Agency CNPq for Grant No. 303284/2021-8 and São Paulo Research Foundation (FAPESP) for Grant No. \#2020/02044-9. This work used resources of the Centro Nacional de Processamento de Alto Desempenho em S\~ao Paulo (CENAPAD-SP). The authors acknowledge the National Laboratory for Scientific Computing (LNCC/MCTI, Brazil) for providing HPC resources of the SDumont supercomputer, contributing to the research results reported within this paper. URL: http://sdumont.lncc.br.

\printcredits
\bibliographystyle{unsrt}
\bibliography{cas-refs}

\end{document}